\documentclass{elsart}
\usepackage{feynmf}
\begin{document}
\begin{fmffile}{fmfpaper}
\input epsf

\begin{frontmatter}

\title{Asymptotic expansion of the lattice scalar 
propagator in coordinate space} 

\author{Beatrice Paladini},
\address{School of Mathematics, 
  Trinity College Dublin, 
  Dublin 2, Ireland}
\address{E-mail address: paladini@maths.tcd.ie}
\author{James C. Sexton}, 
\address{School of Mathematics, 
  Trinity College Dublin, 
  Dublin 2, Ireland}
\address{and}
\address{Hitachi Dublin Laboratory,
  16, Westland Row, 
  Dublin 2, Ireland}
\address{E-mail address: sexton@maths.tcd.ie}

\begin{abstract} 
  The asymptotic expansion of the massive scalar field propagator on a 
$n$- dimensional lattice is derived. The method used is based on the 
evaluation of the asymptotic expansion of the modified Bessel function 
$\mathrm{I}_{\nu}(\nu^{2} \beta)$ as the order $\nu$ grows to infinity.
\end{abstract}

\end{frontmatter}

\section{Introduction}

\indent In perturbation theory, the advantage of a coordinate space 
description as a way of studying the 
divergences arising in Feynman 
diagrams has been discussed by many authors 
({\it e.g.}\ \cite{Col,Freed}). 
In the continuum, the analytic expression for the scalar field
propagator in position space is well-known.
On the lattice, however, the standard representation for
the scalar propagator involves integrals over Bessel functions and 
has proved to be very difficult to analyse in the continuum limit.\\
\indent Several attempts have been made to derive a suitable expansion 
for the lattice scalar propagator in the limit where the lattice spacing
goes to zero. In
particular, we mention the analysis performed by  
L\"{u}scher and Weisz for the massless propagator \cite{Lus} as well as the 
research carried out by Burgio, Caracciolo and Pelissetto in \cite{Pel}. 
In the first of these studies, the 
$x$- dependence of the massless propagator is asymptotically derived. 
In the second, the $m$-dependence is 
obtained for the propagator at $x = 0$. The propagator at any other
point $x \neq 0$ is then expressed, through a set of recursion
relations, in terms of the value which its massless counterpart assumes on  
an element of the unit hypercube (i. e, $x_{\mu}$ either 0 or
1). In the present paper we wish to tackle the most general case of both
$m$ and $x$ non-vanishing. \\  
\indent The procedure that we adopt attacks
the question directly at its core. We derive  
an asymptotic expansion of the modified Bessel function of the first 
kind which appears in the expression of the lattice propagator.  The
order
and argument of this Bessel function go to infinity at different rates
as the lattice spacing decreases to zero.
We have found no tabulated asymptotic expansion for this particular
case. Expansions for the modified Bessel function of the 
first kind, $I_{\nu}(x)$, are, indeed, available for the cases where 
either the argument $x$ or the order $\nu$ becomes large, or for the 
case where both $x$ and $\nu$ grow to infinity keeping the value of 
their respective ratio constant (approximantion by
tangents). Unfortunately, none of the cases just mentioned characterizes
the modified Bessel function at hand, and we have had to develop, 
as a result, a new asymptotic expansion.\\
\noindent As a perturbative application of the technique developed in   
the analysis of the continuum limit expansion of the lattice scalar 
propagator, in the closing section of this paper we consider the 
mass renormalization of the discrete $\lambda \phi^{4}$ theory in 
coordinate space. \\ 
\indent We introduce now briefly the notations that will be assumed in the 
following sections. Throughout this paper, we shall work on a 
$n$-dimensional lattice of finite spacing $a$. The convention  
\begin{equation}
x^{p} = \sum_{\mu = 1}^{n} x_{\mu}^{p}, \; \; \; \; \mbox{with $p$ an 
integer} \; ,    \label{eq:notation} 
\end{equation}
\noindent will also be used. \\
\noindent The Euclidean free scalar propagators in a $n$-dimensional 
configuration space will be denoted by $\Delta^{\mathrm{C}}(x;n)$ and 
$\Delta^{\mathrm{L}}(x; n)$ with
\begin{equation}
\Delta^{\mathrm{C}}(x; n) = \int_{- \infty}^{+ \infty} \frac{d^{n}p}{(2
\pi)^n} \frac{e^{\imath p x}}{m^2 + p^2} \; ,  \label{eq:defcon}
\end{equation}
\begin{equation}
\Delta^{\mathrm{L}}(x; n) = \int_{- \frac{\pi}{a}}^{+ \frac{\pi}{a}} 
\frac{d^{n}p}{(2 \pi)^n} \frac{e^{\imath p x}}{m^2 + \hat{p}^2} 
\; ,  \label{eq:deflat} 
\end{equation}
\noindent referring to the propagator evaluated in the continuum and on 
the lattice, respectively. Note that in eq. (\ref{eq:deflat}) we have 
introduced the short-hand notation 
\begin{equation}
\hat{p}^2 = \frac{4}{a^2} \sum_{\mu = 1}^{n} \sin^{2} \left(
\frac{p_{\mu} a}{2} \right).   \label{eq:shorthand}  
\end{equation}

\section{The propagator on the lattice}

\indent It is a well- known result that the continuum scalar field 
propagator in a 4- dimensional configuration space can be written in 
terms of the modified Bessel function $\mathrm{K}_{1}$. 
The technique consists in
using the Schwinger's representation for the propagator in
momentum space. Re-expressing, then, the integral over the four-momentum
as a product of four one-dimensional integrals and completing the 
square in the resulting exponential, we can finally perform the
integration and obtain our final formula. Adopting the notational
convention given by eq. (\ref{eq:notation}), 
we write the $n$-dimensional expression as follows 
\begin{eqnarray}
\Delta^{\mathrm{C}}(x; n) & = & \int_{- \infty}^{+ \infty} \frac{d^{n}p}{(2
\pi)^n} \frac{e^{\imath p x}}{m^2 + p^2} \nonumber \\  
& = & \int_{0}^{\infty} d\alpha e^{- m^{2} \alpha} \prod_{\mu = 1}^{n} 
\int_{- \infty}^{+ \infty} \frac{dp_{\mu}}{(2 \pi)} \exp \left \{- p_{\mu}^2 
\alpha + \imath p_{\mu} x_{\mu} \right \} \nonumber \nonumber \\ 
& = & (2 \pi)^{- n/2} \left[ \frac{(x^2)^{1/2}}{m} \right]^{1 - n/2}  
\mathrm{K}_{1 - n/2}\left[ m (x^2)^{1/2} \right].    \label{eq:npropcon} 
\end{eqnarray}
\indent The derivation of the standard representation for the 
lattice propagator in configuration space is carried out in a much similar
fashion. Indeed, we have 
\begin{eqnarray}
\Delta^{\mathrm{L}}(x; n) & = & \int_{- \frac{\pi}{a}}^{+ \frac{\pi}{a}} 
\frac{d^{n}p}{(2 \pi)^n} \frac{e^{\imath p x}} 
{m^2 + \hat{p}^2} \nonumber \\
& = & \int_{0}^{\infty} d\alpha 
e^{- m^2 \alpha} \left \{ \prod_{\mu = 1}^{n} 
e^{- \frac{2 \alpha}{a^2}} 
\left( \frac{1}{a} \right) 
\int_{- \pi}^{\pi} \frac{d\vartheta_{\mu}}{(2 \pi)} 
e^{\frac{2 \alpha}{a^2} \cos\vartheta_{\mu}} 
\cos \left[ \left( \frac{x_{\mu}}{a} \right) \vartheta_{\mu} 
\right] \right \} \nonumber \\
& = & \int_{0}^{\infty} d\alpha  
e^{- m^2 \alpha} \left \{\prod_{\mu = 1}^n 
e^{- \frac{2 \alpha}{a^2}} \left(\frac{1}{a}
\right) \mathrm{I}_{\frac{x_{\mu}}{a}}
\left(\frac{2 \alpha}{a^2}\right) \right \}    \label{eq:nproplat}
\end{eqnarray}
\noindent with $\mathrm{I}_{\frac{x_{\mu}}{a}}
\left(\frac{2 \alpha}{a^2}\right)$ corresponding to the modified
Bessel function of the first kind. \\
\indent Unfortunately, the integral appearing in eq. (\ref{eq:nproplat})
cannot be trivially solved. As a consequence, we are
not able in this case to express the propagator in closed form. What we 
really wish to do here, though, is to show that in the continuum limit, 
{\it i.e. \/} when $a \rightarrow 0$, $\Delta^{\mathrm{L}}(x; n)$ is given by 
the sum of $\Delta^{\mathrm{C}}(x; n)$ plus a series of correction terms 
depending on increasing powers of the lattice spacing $a$. \\ 
\indent As already mentioned in the introduction, the most direct way to
proceed in order to achieve our goal is trying to derive the asymptotic 
expansion for the modified Bessel function 
$\mathrm{I}_{\frac{x_{\mu}}{a}}(\frac{2 \alpha}{a^2})$ 
as $a \rightarrow 0$. \\
\noindent The strategy to adopt for the actual derivation of the expansion is
determined by the fact that the order and the argument of
the modified Bessel function become large at different rates. The
standard techniques of {\em global \/} analysis (e.g, the steepest descents
method) are not of much use in this case. As a consequence, we are  
forced to choose here a {\em local \/} analysis approach. This implies
beginning our study by examining the differential equation satisfied by 
the modified Bessel function at hand. With the purpose of determining 
uniquely the asymptotic behaviour of the solution, we shall also impose 
in the end the condition that the series representation derived 
reproduce (through its leading term) the continuum result 
$\Delta^{\mathrm{C}}(x; n)$. \\ 
\indent We first commence by setting
\begin{equation}
\beta \equiv \frac{2 \alpha}{x_{\mu}^{2}} \;\;\; {\mbox and} \;\;\;
\nu \equiv \frac{x_{\mu}}{a} \; \; \; ,  \label{eq:set} 
\end{equation}
\noindent {\it i.e. \/} $\, \mathrm{I}_{x_{\mu}/a}(2 \alpha/ 
a^{2}) \rightarrow \mathrm{I}_{\nu}(\nu^{2} \beta)$. \\ 
\noindent Thus, we wish to
find an expansion for $\mathrm{I}_{\nu}(\nu^{2} \beta)$ as $\nu \rightarrow
\infty$. We observe now that $\mathrm{I}_{\nu}(\nu^{2} \beta)$ satisfies
a differential equation of the form: $x^{2} \frac{d^{2}y}{dx^2} + 
x \frac{dy}{dx} - (x^{2} + \nu^{2}) y = 0$. Hence, performing the
change of variable
\begin{equation}
x \rightarrow  \nu^{2} \beta  \label{eq:changex} 
\end{equation}
\noindent we obtain
\begin{equation}
\frac{\partial^{2} \mathrm{I}_{\nu}}{\partial \beta^2}(\nu^{2} \beta) 
+ \frac{1}{\beta} \frac{\partial 
\mathrm{I}_{\nu}}{\partial \beta}(\nu^{2} \beta) - \nu^{2} \left[
\frac{1}{\beta^2} + \nu^{2} \right] \mathrm{I}_{\nu}(\nu^{2} \beta) = 0.  
\label{eq:besdif}  
\end{equation}
\noindent Eq. (\ref{eq:besdif}) simplifies slightly if we make the 
substitution 
\begin{equation}
\mathrm{I}_{\nu}(\nu^{2} \beta) = 
\frac{\mathrm{C}}{\sqrt{\beta}} \mathrm{Y}(\beta, \nu). 
\label{eq:besiny}  
\end{equation}
\noindent $\mathrm{C}$ represents here the free parameter whose value 
shall be fixed later on according to the prescription made at the
beginning of this paragraph. \\ 
\indent By using eq. (\ref{eq:besiny}), we now get a new differential 
equation for $\mathrm{Y}(\beta, \nu)$, namely 
\begin{equation}
\frac{\partial^{2} \mathrm{Y}(\beta, \nu)}{\partial \beta^2} - 
\left[ \nu^{4} + \frac{ \nu^{2} - \frac{1}{4}}{\beta^{2}} \right] 
\mathrm{Y}(\beta, \nu) = 0.  
\label{eq:ydif}   
\end{equation}
\indent We consider at this stage the limit $\nu \rightarrow \infty$. 
We now need to make some kind of assumption on the form of 
leading term governing the expansion of the solution to
eq. (\ref{eq:ydif}) as $\nu$ becomes large. With this purpose, we 
consider a substitution originally 
suggested by Carlini (1817), Liouville (1837) and Green (1837) and whose
underlying idea is that the controlling factor in the asymptotic 
expansion of a function is usually in the form of an exponential. 
Hence, we assume the asymptotic relation 
\begin{equation}
\mathrm{Y}(\beta, \nu) \sim e^{\nu^{2} \beta + 
\int w(\beta, \nu) d\beta}. 
\label{eq:carlini}   
\end{equation}
\indent Actually, the form which we have used here for 
$\mathrm{Y}(\beta, \nu)$ differs slightly from the standard Carlini's 
substitution since it implicitly assumes that not only the 
leading term, but also all the subsequent terms in the expansion are 
expressible as exponentials. The reason which led us to 
adopt eq. (\ref{eq:carlini}) as a viable option is essentially twofold:
on one hand, it is dictated by the necessity of recovering the
well-known expression for the free scalar propagator in the continuum;
on the other, the idea spurred in analogy to the method adopted in 
Watson \cite{Wat} and originally due to Meissel (1892) in order to
deduce the asymptotic expansion for the Bessel function of the first
kind, $\mathrm{J}_{\nu}(\nu z)$ with $\nu$ large. \\
\indent Carrying out the substitution in the 
differential equation for $\mathrm{Y}(\beta, \nu)$, we end up with a 
first-order inhomogeneous differential equation for the new function 
$w(\beta, \nu)$, {\it i.e. \/} 
\begin{equation} 
\frac{\partial w(\beta, \nu)}{\partial \beta} + 
w^{2}(\beta, \nu) + 
2 \nu^{2} w(\beta, \nu) - \frac{\nu^{2} - \frac{1}{4}}{\beta^{2}} 
\sim 0.  \label{eq:wdif} 
\end{equation}
\noindent Our aim, at this point, is to look for a suitable series 
representation for the solution to this equation. With this purpose, we 
make the following observations. First of all, if we suppose that 
$2 \nu^{2} w(\beta, \nu) - 
\nu^{2}/ \beta^{2} \sim 0$ represents a good 
approximation to eq. (\ref{eq:wdif}) when $\nu \rightarrow \infty$, then
it is straightforward to derive the leading behaviour of 
$w(\beta, \nu)$ as   
$(2 \beta^2)^{-1}$. We note that this gives exactly the continuum
result $\Delta^{\mathrm{C}}(x; n)$ once the overall solution to
eq. (\ref{eq:besdif}) is normalized by fixing the arbitrary parameter 
$\mathrm{C}$ to be equal to $(2 \pi \nu^2)^{- 1/2}$. Secondly, we observe that
all the non-leading terms in the expansion of $w(\beta, \nu)$ should 
feature only even powers of $\nu$, since only even powers of this 
variable appear in eq. (\ref{eq:wdif}). \\ 
\noindent As a result of the remarks just made, we assume that 
$w(\beta, \nu)$ admits an asymptotic series representation for 
$\nu$ large of the type  
\begin{equation}
w(\beta, \nu) = \sum_{n = 0}^{\infty} a_{n}(\beta) \nu^{- 2n}.  
\label{eq:wseries}
\end{equation}
\noindent This is consistent with the approximation that we considered
when we deduced the form of the leading term of the solution. 
Substituting now this formula into eq. (\ref{eq:wdif}), we get 
\begin{equation}
2 \sum_{n = - 1}^{\infty} a_{n + 1}(\beta) \nu^{- 2n} \sim     
\frac{\nu^{2} - \frac{1}{4}}{\beta^{2}} - \sum_{n = 0}^{\infty} 
\left \{ \left[ \sum_{m = 0}^{n} a_{m}(\beta) 
a_{n - m}(\beta) \right] +   
\frac{\partial}{\partial \beta} a_{n}(\beta) \right \} \nu^{-2n}.  
\label{eq:wdifference}   
\end{equation}
\indent Matching, order by order, the coefficients corresponding to the 
same power of $\nu$, we are now able to deduce the set of relations defining
the coefficients $a_{n}(\beta)$. We have
\begin{eqnarray}
a_{0}(\beta) & = & \frac{1}{2 \beta^2} \; \; \; \; \; ;  
\; \; \; \; \; a_{1}(\beta) = - \frac{1}{8 \beta^{2}} 
+ \frac{1}{2 \beta^{3}} - \frac{1}{8 \beta^{4}} \; \; \; \; \; ; 
\label{eq:a0a1} \\
a_{n + 1}(\beta) & = & - \frac{1}{2} 
\left \{ \sum_{m = 0}^{n} a_{m}(\beta) 
a_{n - m}(\beta) + \frac{\partial}{\partial \beta} 
a_{n}(\beta) \right
\} \; \; \; \; \; n \geq 1.   \label{eq:anrec} 
\end{eqnarray}
\noindent The coefficients $a_{n}(\beta)$ can be, at this stage,
computed iteratively leading to 
\begin{eqnarray}
\lefteqn{w(\beta, \nu) = 
\frac{1}{2 \beta^{2}} - \left[ \frac{1}{8 \beta^{2}}
- \frac{1}{2 \beta^{3}} + \frac{1}{8 \beta^{4}} \right] \nu^{- 2}} 
\nonumber \\  
& & + \left[  - \frac{1}{8 \beta^{3}} +
\frac{13}{16 \beta^{4}} - \frac{1}{2 \beta^{5}} + \frac{1}{16 \beta^{6}}
\right] \nu^{- 4} \nonumber \\ 
& & + \left[ - \frac{25}{128 \beta^{4}} +  
\frac{7}{4 \beta^{5}} - \frac{115}{64 \beta^{6}} + \frac{1}{2 \beta^{7}}
- \frac{5}{128 \beta^{8}} \right] \nu^{- 6} + {\mathrm{O}}(\nu^{- 8}).  
\label{eq:wexp}   
\end{eqnarray}
\indent The asymptotic expansion of $\mathrm{I}_{\nu}(\nu^{2}
\beta)$ as $\nu \rightarrow \infty$ can be now easily derived. The result 
reads as follows 
\begin{eqnarray}
\lefteqn{\mathrm{I}_{\nu}(\nu^{2} \beta) \sim _{\nu \rightarrow \infty}   
(2 \pi \nu^{2} \beta)^{- \frac{1}{2}} 
\exp \left( \nu^{2} \beta - \frac{1}{2 \beta} \right)} \nonumber \\ 
& & \times \left \{ 1 + \left[ \frac{1}{8 \beta} - 
\frac{1}{4 \beta^2} + \frac{1}{24 \beta^3} \right] \nu^{- 2} \nonumber
\right. \\ 
& & \; \; \; \; \; + \left[ \frac{9}{128 \beta^2} - 
\frac{29}{96 \beta^3} + \frac{31}{192 \beta^4} - 
\frac{11}{480 \beta^5} + \frac{1}{1152 \beta^6} \right] \nu^{- 4}  
\nonumber \\ 
& & \; \; \; \; \; + \left[ \frac{75}{1024 \beta^3} - 
\frac{751}{1536 \beta^4} + \frac{1381}{3072 \beta^5} 
- \frac{1513}{11520 \beta^6} + \frac{4943}{322560 \beta^7} \right.  
\nonumber \\  
& & \; \; \; \; \; \; \; - \left. \left. \frac{17}{23040 \beta^8} + 
\frac{1}{82944 \beta^9} \right] \nu^{- 6} + 
{\mathrm{O}}(\nu^{- 8})  \right \}.   \label{eq:besexp} 
\end{eqnarray}
\noindent Dividing this equation by $\mathrm{I}_{\nu}(\nu^2 \beta)$ 
and plotting the
result against $\beta$ for a fixed (large) value of $\nu$, we see
a fluctuation of the ratio around unity. This reproduces exactly the
behaviour which ought to be expected by the ratio of two functions 
asymptotic to each other. Furthermore, we observe that the 
fluctuation around 1 is a feature of the whole positive $\beta$-axis, 
therefore implicitly suggesting the validity of our expansion even for 
large values of $\beta$. 

\begin{figure}[h]
\epsfxsize = 10cm
\epsfbox{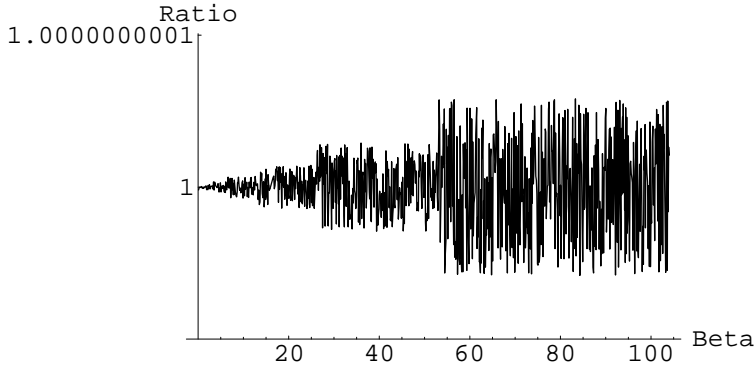}
\caption {Behaviour of the ratio of the asymptotic expansion of 
$\mathrm{I}_{\nu}(\nu^2 \beta)$ with $\mathrm{I}_{\nu}(\nu^2 \beta)$ 
as a function of $\beta$ with $\nu$ equal to 100.}   \label{fig:ratio} 
\end{figure}

\indent Recalling the expressions of $\beta$ and $\nu$ in terms of 
$\alpha$, $x_{\mu}$ and $a$, we are now able to derive the formula
for the $n$- dimensional propagator up to sixth order in the lattice 
spacing. In fact, we can do more than that. We consider at this point a 
generalisation of $\Delta^{\mathrm{L}}(x; n)$ by introducing an arbitrary
exponent $q$ in the denominator of eq. (\ref{eq:nproplat}); 
that is, we now look at  
\begin{eqnarray}
\Delta^{\mathrm{L}}(x; n; q) & = & \int_{- \frac{\pi}{a}}^{+ \frac{\pi}{a}} 
\frac{d^{n}p}{(2 \pi)^n} \frac{e^{\imath p x}} 
{\left \{ m^2 + \hat{p}^2 \right \}^{q}} \nonumber \\
& = & \frac{1}{\Gamma(q)} \int_{0}^{\infty} d\alpha \alpha^{q - 1}   
e^{- m^2 \alpha} \left \{\prod_{\mu = 1}^n 
e^{- \frac{2 \alpha}{a^2}} \left(\frac{1}{a}
\right) \mathrm{I}_{\frac{x_{\mu}}{a}}
\left(\frac{2 \alpha}{a^2}\right) \right \}.   \label{eq:general}   
\end{eqnarray}
\noindent The main reason for considering this generalised 
quantity rests in the fact that eq. (\ref{eq:general}) represents a 
key element in the expression of general one-loop lattice integrals 
with bosonic propagators and zero external momenta \cite{Pel}. \\
\indent Having reached a formula for the asymptotic expansion of the 
modified Bessel function as the lattice spacing vanishes, the study of the 
continuum limit of $\Delta^{\mathrm{L}}(x; n; q)$ is not technically
more difficult than the analysis of the same limit for the
$n$- dimensional lattice propagator $\Delta^{\mathrm{L}}(x; n) \equiv 
\Delta^{\mathrm{L}}(x; n; 1)$. Indeed, we simply have now to substitute eq. 
(\ref{eq:besexp}) into eq. (\ref{eq:general}) and carry out the product 
over the dimensional index $\mu$. 
The resulting $\alpha$-integrals are all well-defined and finite. We
can, therefore, proceed to their evaluation and obtain 
\begin{eqnarray}
\Delta^{\mathrm{L}}(x; n; q) & \sim_{a \rightarrow 0} & 
\frac{(4 \pi)^{- n/2}}{\Gamma(q)} \left \{
\Delta^{\mathrm{L}}_{0}(x; n; q) + 
a^2 \Delta^{\mathrm{L}}_{2}(x; n; q) \right. \nonumber \\
& & + \left. a^4 \Delta^{\mathrm{L}}_{4}(x; n; q) + 
a^6 \Delta^{\mathrm{L}}_{6}(x; n; q)
+ {\mathrm{O}} (a^8) \right \}.   \label{eq:genexp}
\end{eqnarray}
\noindent The full expression of each of the coefficients 
$\Delta^{\mathrm{L}}_{2i}(x; n; q)$ 
($i = 0, 1, 2, 3$) is given in terms of the  
new function ${\mathrm{P}}_{\rho}(m; x)$ defined as 
\begin{equation}
{\mathrm{P}}_{\rho}(m; x) = \left[ \frac{2 m}{(x^2)^{\frac{1}{2}}}
\right]^{\rho} \mathrm{K}_{\rho} \left[ m \left( x^2 \right)^{\frac{1}{2}}
\right]   \label{eq:deffunp} 
\end{equation}
\noindent with $\mathrm{K}_{\rho} 
\left[ m \left( x^2 \right)^{\frac{1}{2}} \right]$ 
representing, as usual, the modified Bessel function of the second kind
and $\rho$ a real number. Therefore, we have  
\begin{eqnarray}
\lefteqn{\Delta^{\mathrm{L}}_{0}(x; n; q) = 
2 {\mathrm{P}}_{\frac{n}{2} - q}(m; x)} 
\label{eq:zeroord} \\
\lefteqn{\Delta^{\mathrm{L}}_{2}(x; n; q) =  
\frac{n}{8} {\mathrm{P}}_{1 + \frac{n}{2} - q}(m; x) -   
\frac{x^2}{8} {\mathrm{P}}_{2 + \frac{n}{2} - q}(m; x) +   
\frac{x^4}{96} {\mathrm{P}}_{3 + \frac{n}{2} - q}(m; x)}  
\label{eq:secondord} \\
\lefteqn{\Delta^{\mathrm{L}}_{4}(x; n; q) =  
\frac{n (n + 8)}{256} {\mathrm{P}}_{2 + \frac{n}{2} - q}(m; x) -   
\left[ \left( \frac{n}{128} + \frac{13}{192} \right) x^2\right] 
{\mathrm{P}}_{3 + \frac{n}{2} - q}(m; x)} \nonumber \\  
& + & \left[\frac{n + 24}{1536} x^4 + \frac{(x^2)^2}{256} \right] 
{\mathrm{P}}_{4 + \frac{n}{2} - q}(m; x) \nonumber \\ 
& - & \left[\frac{x^6}{1280} + \frac{x^2 x^4}{1536} \right] 
{\mathrm{P}}_{5 + \frac{n}{2} - q}(m; x) + 
\frac{(x^4)^2}{36864} {\mathrm{P}}_{6 + \frac{n}{2} - q}(m; x)  
\label{eq:fourthord} \\ 
\lefteqn{\Delta^{\mathrm{L}}_{6}(x; n; q) =  
\left[ \frac{n (n - 1) (n - 2)}{12288} + 
\frac{3 n (3 n + 22)}{4096} \right] 
{\mathrm{P}}_{3 + \frac{n}{2} - q}(m; x)} \nonumber \\ 
& - & \left[ \frac{(n - 1) (n - 2)}{4096} + 
\frac{85 n + 666}{12288} \right] x^2     
{\mathrm{P}}_{4 + \frac{n}{2} - q}(m; x) \nonumber \\  
& + & \left[ \frac{(n - 1) (n - 2) + 59 n + 1102}{49152} x^4 + 
\frac{3 n + 52}{12288} (x^2)^2 \right] 
{\mathrm{P}}_{5 + \frac{n}{2} - q}(m; x) \nonumber \\   
& - & \left[\frac{3 n + 160}{61440} x^6 + \frac{(x^2)^3}{12288} + 
\frac{3 n + 98}{73728} x^2 x^4 \right] 
{\mathrm{P}}_{6 + \frac{n}{2} - q}(m; x) \nonumber \\ 
& + & \left[\frac{5}{57344} x^8 + \frac{x^4 (x^2)^2}{49152} + 
\frac{n + 48}{589824} (x^4)^2 + \frac{x^2 x^6}{20480} \right] 
{\mathrm{P}}_{7 + \frac{n}{2} - q}(m; x) \nonumber \\
& - & \left[\frac{x^2 (x^4)^2}{589824} + \frac{x^4 x^6}{245760} \right] 
{\mathrm{P}}_{8 + \frac{n}{2} - q}(m; x) + 
\frac{(x^4)^3}{21233664} {\mathrm{P}}_{9 + \frac{n}{2} - q}(m; x). 
\label{eq:sixthord}    
\end{eqnarray}
\indent The expansion obtained for $\Delta^{\mathrm{L}}(x; n; q)$ 
clearly shows how the 
finite corrections introduced by formulating the theory on a lattice can
be analytically expressed by a series of increasing (even) powers of 
the lattice spacing {\em a \/} with coefficients given by analytic
functions of the mass and space coordinate times a modified Bessel 
function of the second kind of increasing order $\rho$. \\ 
\indent We intend now to demonstrate how eq. (\ref{eq:genexp}) is in 
perfect agreement with both the studies performed in 
\cite{Lus} and \cite{Pel}. 
With this aim, we analyse $\Delta^{\mathrm{L}}(x; n; q)$ in the limit 
$m (x^2)^{1/2} \rightarrow0$. Given the functional dependence of eq. 
(\ref{eq:genexp}) on ${\mathrm{P}}_{\rho}(m; x)$ and given the definition 
in eq. (\ref{eq:deffunp}), this translates into considering the 
appropriate expansion for the Bessel function 
$\mathrm{K}_{\rho}\left[ m (x^2)^{\frac{1}{2}} \right]$. 
We wish to recall at this point that the series representation of the 
modified Bessel function of the second kind assumes different forms 
depending on whether the order $\rho$ is a real ($\rho_{\mathrm{re}}$) 
or integer ($\rho_{\mathrm{in}}$). 
In particular, for $m (x^2)^{1/2}$ small and 
$\rho = \rho_{\mathrm{re}}$ we have
\begin{equation}
{\mathrm{P}}_{\rho_{\mathrm{re}}}(m; x) \sim  
\frac{\pi}{2} \frac{1}{\sin \rho_{\mathrm{re}} \pi} 
\left \{ \frac{2^{2 \rho_{\mathrm{re}}}}
{\Gamma(1 - \rho_{\mathrm{re}})} \frac{1}{(x^2)^{\rho_{\mathrm{re}}}} - 
\frac{m^{2 \rho_{\mathrm{re}}}}
{\Gamma(1 + \rho_{\mathrm{re}})} \right \}   \label{eq:preal}  
\end{equation}
\noindent while, for $m (x^2)^{1/2}$ still small and $\rho =
\rho_{\mathrm{in}}$ ($\rho_{\mathrm{in}} \neq 0$) 
\footnote{For $\rho_{\mathrm{in}} = 0$ the 
correct expansion reads ${\mathrm{P}}_{0}(m ; x) \sim 
\psi(1) - \ln \left[ \frac{m (x^2)^{1/2}}{2} \right]$.}, we find
\begin{eqnarray}
\lefteqn{{\mathrm{P}}_{\rho_{\mathrm{in}}}(m; x) \sim  
\frac{2^{2 \rho_{\mathrm{in}} - 1}}{(x^2)^{\rho_{\mathrm{in}}}}
\Gamma(\rho_{\mathrm{in}})} \nonumber  \\ 
& + & (- 1)^{\rho_{\mathrm{in}} + 1} 
\frac{m^{2 \rho_{\mathrm{in}}}}{\Gamma(1 + \rho_{\mathrm{in}})} 
\left \{ \ln \left[\frac{m (x^2)^{1/2}}{2} \right] 
- \frac{1}{2} \psi(1) - \frac{1}{2} \psi(1 + \rho_{\mathrm{in}}) \right \}
\label{eq:pinteger} 
\end{eqnarray}
\noindent with $\psi$ denoting the $\psi$- function \cite{Ryz}. \\  
\noindent Using the relation $\Gamma(1 - \rho_{\mathrm{re}}) 
\Gamma(\rho_{\mathrm{re}}) = \pi/ \sin \rho_{\mathrm{re}} \pi$, we  find that  
${\mathrm{P}}_{\rho_{\mathrm{re}}}(m; x)$ and 
${\mathrm{P}}_{\rho_{\mathrm{in}}}(m; x)$ assume    
the same functional form $2^{\rho - 1} \Gamma(\rho)/(x^2)^{\rho}$ 
($\rho > 0$) as $m \rightarrow 0$. The limits $m \rightarrow 0$ and 
$\rho_{\mathrm{re}} \rightarrow \rho_{\mathrm{in}}$ are, therefore,
uniform. After performing the final mechanical analysis to set $n=4$ and
$q=1$, we find that the massless limit of our continuum 
expansion for $\Delta^{\mathrm{L}}(x; n; q)$ reproduces exactly the 
expansion obtained by 
L\"{u}scher and Weisz for the massless 4-dimensional propagator \cite{Lus}\\
\noindent The limit $x \rightarrow 0$ proves to be more difficult 
to analyse. The short-distance behaviour is, indeed, singular and the 
limit not uniform in this case. Note that this observation matches the 
analogous remark made in \cite{Col} about the behaviour of the dimensionally 
regularised propagator. Observe also that the logarithmic
mass-behaviour described in \cite{Pel} by Burgio, Caracciolo and
Pelissetto is recovered, in our formulation, for integral values of
$\rho$.

\section{The tadpole diagram in $\lambda \phi^{4}$ theory}

\indent As a direct implementation of the result obtained for 
the continuum expansion of the scalar propagator, we now 
wish to derive the one-loop renormalization mass counterterm of 
$\lambda \phi^{4}$ theory through the study of the lattice 
tadpole diagram.  

\begin{figure}[h]
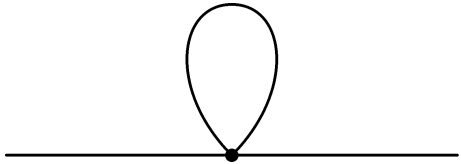

\begin{center}
\begin{fmfchar*}(40,25)
 \fmfleft{i} 
 \fmfv{label=$x$, label.angle=-90,label.dist=50}{i}
 \fmfright{o} 
 \fmfv{label=$y$, label.angle=-90,label.dist=50}{o}
 \fmf{plain}{i,v,v,o}
 \fmfdot{v}
 \fmfv{label=$z$,label.angle=-90,label.dist=50}{v} 
\end{fmfchar*}  \label{fig:tadpole}
\end{center}
\caption{Self-energy tadpole diagram in $\lambda \phi^{4}$ theory.} 
\end{figure}

\noindent In a $n$-dimensional continuum space, the contribution to the 
full propagator associated with the graph in Fig. 2 is well-known and 
proportional to the integral  
$\int d^{n}z \Delta^{\mathrm{C}}(x - z) \Delta^{\mathrm{C}}(0) 
\Delta^{\mathrm{C}}(z - y)$. The 
ultraviolet behaviour of the diagram is entirely due to the divergence 
in $\Delta^{\mathrm{C}}(0)$ \cite{Col}. \\
\indent In terms of our notational conventions, the lattice version of 
the tadpole graph in $n$-dimensions is immediately written as
\begin{equation}
\mathrm{M}^{\mathrm{L}}_{\mathrm{tad}} = - \mu^{4 - n} 
\left(\frac{\lambda}{2}\right) a^{n} \sum_{z} 
\Delta^{\mathrm{L}}(x - z; n) \Delta^{\mathrm{L}}(z - y; n) 
\Delta^{\mathrm{L}}(0; n) \; \; ,  \label{eq:tadlat}
\end{equation}
\noindent with $\lambda$ the coupling constant and $\mu^{4 - n}$ a 
multiplicative dimensional factor introduced to preserve the dimensional
correctness of the theory. Note that, in eq. (\ref{eq:tadlat}), the lattice
propagators have replaced their continuous counterparts and the
summation over the lattice sites has taken the place of the continuum 
integral. \\
\indent Our present goal is to examine eq. (\ref{eq:tadlat}) as $a
\rightarrow 0$. Using the asymptotic expansion of the propagator as 
derived in eq. (\ref{eq:genexp}), it is straightforward to 
obtain up to fourth order in the lattice spacing  
\begin{eqnarray}
& & \mathrm{M}^{\mathrm{L}}_{\mathrm{tad}} \sim \int d^{n}z 
\left \{ f(x, y, z) \Delta^{\mathrm{L}}_{0}(0; n) + 
a^2 \left[ g(x, y, z) \Delta^{\mathrm{L}}_{0}(0; n) + 
f(x, y, z) \Delta^{\mathrm{L}}_{2}(0; n) \right] + \right. \nonumber \\  
& & \; \; \; \; \; \; 
\left. a^4 \left[ h(x, y, z) \Delta^{\mathrm{L}}_{0}(0; n) + 
g(x, y, z) \Delta^{\mathrm{L}}_{2}(0; n) + 
f(x, y, z) \Delta^{\mathrm{L}}_{4}(0; n) \right]  
+ \ldots \right \} \; ,  \label{eq:tadexp}
\end{eqnarray}
\noindent with $f(x, y, z)$, $g(x, y, z)$ and $h(x, y, z)$
denoting functions which are associated with products 
of the type $\Delta^{\mathrm{L}}_{0} \Delta^{\mathrm{L}}_{0}$, 
$\Delta^{\mathrm{L}}_{2} \Delta^{\mathrm{L}}_{0}$ 
and $\Delta^{\mathrm{L}}_{4} \Delta^{\mathrm{L}}_{0}$ plus 
$\Delta^{\mathrm{L}}_{2} \Delta^{\mathrm{L}}_{2}$,
respectively. \\
\noindent We observe that the n-dimensional coefficients which 
appear in the expressions of $f$, $g$ and $h$ are exclusively evaluated at 
$x - z$ and $z - y$ and correspond, hence, to infinities which are, 
ultimately, integrable. As a result, the singularities in 
$\mathrm{M}^{\mathrm{L}}_{\mathrm{tad}}$ are fundamentally generated
by the poles in $\Delta^{\mathrm{L}}_{0}(0; n)$, 
$\Delta^{\mathrm{L}}_{2}(0; n)$ and 
$\Delta^{\mathrm{L}}_{4}(0; n)$ only. It is now of paramount importance 
to remark that each of the latter quantities scales, implicitly, 
with the lattice spacing $a$. This scaling needs to be made explicit in
the analysis by considering the mass as physical, 
{\it i.e.\/} $m_{\mathrm{R}} = m a$, and by recalling that, in a discrete
formulation, the space coordinate is also defined in terms of the
lattice spacing. \\
\noindent At this stage, the isolation of the $\mathrm{UV}$-divergences 
can take place along the lines of the method introduced in Collins for 
the study of the continuum case. Indeed, performing a {\em point-splitting\/} 
of the tadpole $z$-vertex through the introduction of an arbitrary 
variable $\varepsilon_{\mathrm{R}} = \varepsilon / a$, such 
that $\varepsilon_{\mathrm{R}} \rightarrow 0$ (a fixed), the 
investigation of the divergent behaviour in each of the 
$\Delta^{\mathrm{L}}_{2 i}(0; n)$ ($i = 0, 1, 2$) translates now into 
examining the divergences in  
$\Delta^{\mathrm{L}}_{2 i}(\varepsilon_{\mathrm{R}}; m_{\mathrm{R}}; n)$
\footnote{The explicit presence of the mass in the argument of the propagator 
coefficients has been included, in this case, simply to recall that it is 
$m_{\mathrm{R}}$ rather than the bare $m$ which now figures in 
the calculation.} for vanishing $\varepsilon_{\mathrm{R}}$ and $n$ fixed
and non-integral. We note that each of the series expansions obtained 
splits, naturally, into two sub-series, with the poles in 
$\varepsilon_{\mathrm{R}}$ all contained in the first. In terms of 
$\varepsilon_{\mathrm{R}}$, the second sub-series is, in fact,
analytic. Thus, focusing on the singular contribution, we take the 
limit $n \rightarrow 4$ and extract the infinities in 
$\varepsilon_{\mathrm{R}}$ through the singularities of the 
$\Gamma$-function which appears in the series. The procedure leads to a 
renormalization mass counterterm of the type 
\begin{eqnarray}
& & \delta_{\mathrm{L}}m^{2}_{\mathrm{R}} = \frac{1}{a^2} \left \{  
\frac{m^{2}_{\mathrm{R}}}{8 \pi^{2}} \frac{\mu^{(n - 4)}}{(n - 4)} \right \} + 
a^{2} \left \{ \left(\frac{1}{a^{2}}\right)^{2} 
\left[ \frac{m^{2}_{\mathrm{R}}}{8 \pi^{2}} \frac{\mu^{(n - 4)}}{(n - 4)} -  
\frac{m^{4}_{\mathrm{R}}}{64 \pi^{2}}  \frac{\mu^{(n - 4)}}{(n - 4)} \right] + 
\right \} \nonumber \\
& & a^{4} \left \{ \left(\frac{1}{a^{2}}\right)^{3} 
 \left[\frac{m^{2}_{\mathrm{R}}}{8 \pi^{2}} \frac{\mu^{(n - 4)}}{(n - 4)} - 
\frac{m^{4}_{\mathrm{R}}}{64 \pi^{2}}  \frac{\mu^{(n - 4)}}{(n - 4)} + 
\frac{m^{6}_{\mathrm{R}}}{1536 \pi^{2}} \frac{\mu^{(n - 4)}}{(n -
4)}\right] \right\} + \ldots . \label{eq:countn}
\end{eqnarray}
\noindent As expected, the counterterm evaluated through the study of
the lattice continuum limit in $n$-dimensions contains a divergence in
$n$ for $n \rightarrow 4$ as well as a quadratic divergence for 
$a \rightarrow 0$. However, multiplying eq. (\ref{eq:countn}) 
by $a^{2}$, the second order pole in $a$ disappears completely leaving 
the equation, now expressed in lattice units, finite (for $n \neq 4$). \\
\indent The lattice derivation of the mass conterterm can be also performed 
directly in four-dimensions. Due to the important differences existing 
between the expansions in eq. (\ref{eq:preal}) and
eq. (\ref{eq:pinteger}), a few changes take now place in the
investigation, though. The isolation of the infinities is no longer 
accomplished through the extraction of the poles in the
$\Gamma$-functions, but resides ultimately in the observation that the 
divergences in the lattice spacing mirror exactly the analogue divergent
behaviours in $\varepsilon_{\mathrm{R}}$. Operationally, the latter
remark produces a four-dimensional lattice counterterm of the type  
\begin{eqnarray}
& & \delta^{4}_{\mathrm{L}}m^{2}_{\mathrm{R}} =  
\frac{1}{4 \pi^{2}} \left \{ \left(\frac{1}{a^{2}}\right) +   
\frac{m^{2}_{\mathrm{R}}}{4 a^{2}} \log(m^{2}_{\mathrm{R}}) - 
\frac{m^{2}_{\mathrm{R}}}{4 a^{2}} \left[ 4 + \psi(1) + \psi(2)\right] + 
\right. \nonumber \\
& & a^{2} \left \{ \left(\frac{1}{a^{2}}\right) +   
\frac{m^{2}_{\mathrm{R}}}{4 a^{2}} \log(m^{2}_{\mathrm{R}}) - 
\frac{m^{2}_{\mathrm{R}}}{4 a^{2}} \left[ 4 + \psi(1) + \psi(2)\right] + 
\right. \nonumber \\
& & \; \; \left. \left(\frac{1}{4 \pi}\right) \left \{ 
\frac{4}{(a^{2})^{2}} - \frac{1}{8} 
\left(\frac{m^{2}_{\mathrm{R}}}{a^{2}}\right)^{2} 
\left[ \log\left(\frac{m^{2}_{\mathrm{R}}}{4}\right) - \psi(1) - \psi(3)
\right] \right \} \right \} + \nonumber \\
& & a^{4} \left \{ \left(\frac{1}{a^{2}}\right) +   
\frac{m^{2}_{\mathrm{R}}}{4 a^{2}} \log(m^{2}_{\mathrm{R}}) - 
\frac{m^{2}_{\mathrm{R}}}{4 a^{2}} \left[ 4 + \psi(1) + \psi(2)\right] + 
\right. \nonumber \\
& & \; \; \left(\frac{1}{4 \pi}\right) \left \{ 
\frac{4}{(a^{2})^{2}} - \frac{1}{8} 
\left(\frac{m^{2}_{\mathrm{R}}}{a^{2}}\right)^{2} 
\left[ \log\left(\frac{m^{2}_{\mathrm{R}}}{4}\right) - \psi(1) - \psi(3)
\right] \right \} + \nonumber \\
& & \; \; \left. \left. \left(\frac{1}{4 \pi}\right) \left \{ 
\frac{12}{(a^{2})^{3}} + \frac{3}{64} 
\left(\frac{m^{2}_{\mathrm{R}}}{a^{2}}\right)^{3} 
\left[ \log\left(\frac{m^{2}_{\mathrm{R}}}{4}\right) - \psi(1) - \psi(4)
\right] \right \} \right \} + \ldots  \right\} \label{eq:count4}
\end{eqnarray}
\noindent We notice that, in agreement with standard results \cite{Mun},
in this case both a quadratic and a logarithmic counterterm is 
needed at every order in the calculation. Nevertheless, in terms of 
lattice units, eq.(\ref{eq:count4}) is, again, finite.

\section{Conclusions} 

\indent In the present work, we derived an asymptotic expansion for 
the modified Bessel function $\mathrm{I}_{\nu}(\nu^{2} \beta)$ as $\nu 
\rightarrow \infty$. The expansion obtained was of vital importance to 
analytically evaluate the continuum expansion of both the lattice 
scalar propagator in a $n$-dimensional configuration space and its 
related generalised quantity $\Delta^{\mathrm{L}}(x; n; q)$. 
The study of the small 
$m (x^2)^{1/2}$-behaviour of $\Delta^{\mathrm{L}}(x; n; q)$ in the limit 
$a \rightarrow 0$ was shown to involve only the standard 
series expansion of modified Bessel functions of either real 
($\rho_{\mathrm{re}}$) or integer ($\rho_{\mathrm{in}}$) order. 
The uniformity of the limits $m \rightarrow 0$ and 
$\rho_{\mathrm{re}} \rightarrow \rho_{\mathrm{in}}$ was 
observed and the L\"{u}scher and Weisz expansion for the massless 
propagator in four-dimensions recovered. The result obtained in \cite{Pel} 
for the mass-dependence was also reproduced for integral values of
$\rho$. Finally, as a perturbative application of the results obtained, 
the one-loop mass counterterm in $\lambda \phi^{4}$ lattice theory was 
evaluated for both the $n-$ and $4-$dimensional case.

\ack{Beatrice Paladini wishes to thank both the European Human and 
Capital Mobility Program and Hitachi Dublin Laboratory for their 
financial support towards the completion of this work.}

\end{fmffile} 
\end{document}